\documentstyle[preprint,prabib,aps,12pt]{revtex}

\begin{document}

\title{Spherically Symmetric Solution for 
Torsion and the Dirac Equation in 5D Spacetime}

\author{Dzhunushaliev V.D. \thanks{
E-mail address: dzhun@freenet.bishkek.su}}
\address{Deprt. of Phys., VCU, Richmond, VA 23284-2000, USA \\
and 
Theor. Phys. Dept., KSNU, Bishkek, 720024, Kyrgyzstan}
\maketitle

\begin{abstract}
Torsion in a 5D spacetime is considered. 
In this case gravitation is defined by the 5D 
metric and the torsion. It is conjectured that 
torsion is connected with a spinor field. 
In this case Dirac's equation becomes the nonlinear 
Heisenberg equation. It is shown that this 
equation has a discrete spectrum of solutions 
with each solution being regular on the whole 
space and having finite energy. 
Every solution is concentrated on the Planck region 
and hence we can say that torsion should 
play an important role in quantum gravity 
in the formation of bubbles of spacetime 
foam. On the basis of the algebraic relation between 
torsion and the classical spinor field in Einstein-Cartan 
gravity the geometrical interpretation of the spinor 
field is considered as ``the square root'' of torsion. 

\end{abstract}

\pacs{}

\section{Introduction}

It is well known \cite{rod}, \cite{sab}, \cite {ham} that the 
inclusion of torsion leads to the result that in the 
Dirac equation there appears a nonlinear term. 
Heisenberg was the first researcher to investigate 
such a nonlinear equation \cite{hz1}, \cite{hz2}. 
He assumed that a quantization of this nonlinear 
equation would give the mass spectrum 
of elementary particles. This attempt 
was not successful. At present it is understood 
that the Heisenberg equation with 
the specific nonlinear term, 
$(\overline{\psi }\gamma ^\mu \gamma ^5\psi )
\gamma _\mu \gamma ^5\psi$, 
is the consequence of including torsion in the 
geometry of spacetime. On the 
other hand it is well known that the 4D Dirac equation 
can be naturally 
generalized to a 5D equation. In this paper 
the Dirac equation is considered 
in 5D spacetime with torsion. This 5D Heisenberg 
equation has a discrete 
spectrum of regular solutions with finite energy. 
Since the kernel of 
this solution is concentrated in a region 
whose size is 
comparable with the Planck length we 
can postulate that torsion plays important role 
in the formation of spacetime foam in quantum gravity. 

\section{5D Heisenberg equation}

In this paper the 5D spacetime is the principal bundle 
with the $U(1)$ 
structural group and Einstein's 4D spacetime as the 
base of this bundle \cite{dzh}. This means that 5D 
spacetime $M^5$ (total space of the bundle) 
can be presented locally as $M^5=M^4 \times U(1)$ 
where $M^4$ is the Einstein's spacetime (base of the 
bundle) and $U(1)$ is the structural group of the bundle 
(fibre of the bundle). The total space $M^5$ is symmetric 
under the action of the structural group $U(1)$ (gauge group) 
and $M^4=M^5/U(1)$. In this case the extra dimension is 
the symmetrical space (gauge group) and the 5D metric 
on the total space has the following form: 
\begin{equation} 
ds^2 = e^{2\phi (x^\alpha)} \left (dx^5 - 
A_\mu (x^\alpha)dx^\mu \right )^2 
+ g_{\mu\nu} (x^\alpha)dx^\mu dx^\nu, 
\label{1a} 
\end{equation} 
here $g_{\mu\nu}$ is the 4D metric on the base of the bundle; 
$\alpha ,\mu ,\nu = 0,1,2,3$ are the spacetime indices; 
$A_\mu$ is the electromagnetic 
potential according to the following theorem 
\cite{sal}, \cite{per}: 
\par
Let $G$ be the group 
fibre of the principal  bundle.  Then  there  is a one-to-one
correspondence between the $G$-invariant metrics on the  total  
space ${\cal X}$
and the triples $(g_{\mu \nu }, A^{a}_{\mu }, h\gamma _{ab})$. 
Here $g_{\mu \nu }$ is Einstein's pseudo  -
Riemannian metric on the base; $A^{a}_{\mu }$ is the gauge field 
of the group $G$ ( the nondiagonal components of 
the multidimensional metric); $h\gamma _{ab}$  is the 
symmetric metric on the fibre. 
\par 
In this case any physical fields will not depend on 
the 5$^{th}$ coordinate introduced on the 5$^{th}$ dimension. 
The Lagrangian for the Dirac spinor 
field in this spacetime 
with gravity and torsion can be 
written in the following manner: 
\begin{equation}
L=\frac{\hbar c}2\left[ i\overline{\psi }
\gamma ^A(\nabla _A\psi )+\frac{mc} 
\hbar \overline{\psi }\psi +(Hermitian-conjugate)\right] +
\frac 1{2k} R,  
\label{1} 
\end{equation} 
here $A,B,C=0,1,2,3,4$ 
are 5D spacetime indices on the total space of bundle; 
$\gamma ^A$ are Dirac matrics  
satisfying the following condition 
$\{\gamma ^A,\gamma ^B\}=2G^{AB}$; $G^{AB}$ 
is the 5D spacetime metric; $k=8\pi G/c^4$; 
$R$ is the 5D Ricci scalar of the affine connection 
$\Gamma ^A_{\bullet BC}$ (all definitions for Riemann-Cartan 
geometry given in Appendix A). 
The covariant derivative of the spinor 
field is defined in the following way \cite{sab}, \cite{hehl}: 
\begin{equation} 
\nabla _A\psi =\left( \partial _A-\frac{1}{4}
\omega _{abA}\gamma ^{[a}\gamma 
^{b]}-\frac 14S_{abA}\gamma ^{[a}\gamma ^{b]}\right) \psi ,
\label{2}
\end{equation}
here $a,b=0,1,2,3,4$ are five-bein indices; 
$\gamma ^a = \gamma ^0, \gamma ^1, \gamma ^2, \gamma ^3, 
\gamma ^4 = \gamma ^5$ are ordinary Dirac
matrices $\{\gamma ^a,\gamma ^b\}=2\eta ^{ab}$; 
$\eta ^{ab} = diag\{ 1,-1,-1,-1,-1\}$ 
is the 5D Minkowski
metric; [ ] means antisymmetrization; \{ \}  
is symmetrization. The coefficients of the spinor 
connection are defined as follows: 
\begin{equation} 
\omega _{abA} = h_{aB}h_b^{\bullet C} 
\left \{ \,^B_{AC}\right \} + 
h_a^{\bullet B}\frac{\partial h_{bB}}{\partial x^A}    
\label{3} 
\end{equation} 
here $h_{\bullet A}^a$ is a five-bein, 
$\left \{ \,^B_{AC}\right \}$ Christoffel symbols. 
We note that for the spinor field defined on the total space 
of principal bundle the covariant derivative with 
respect to extra coordinate (along to fibre of principal 
bundle) is: 
\begin{equation} 
\nabla _5\psi =-\frac{1}{4}\left(\omega _{abA}\gamma ^{[a}\gamma 
^{b]} + S_{abA}\gamma ^{[a}\gamma ^{b]}\right) \psi ,
\label{2a}
\end{equation}
because $\partial _5\psi = 0$ as was indicated above. 
For the totally antisymmetric torsion 
varying the torsion, spinor fields and metric 
leads to the following fields equations: 
\begin{eqnarray}
S^{abc} = -4il_{Pl}^2\left (\overline{\psi }
\gamma ^{[a}\gamma ^b\gamma ^{c]}\psi \right ), &&
\label{4} \\
\left( i\gamma ^A\partial _A-\frac{i}{4}
\omega _{abA}\gamma ^A\gamma
^{[a}\gamma ^{b]} - l_{Pl}^2(\overline{\psi }
\gamma ^{[a}\gamma ^b\gamma
^{c]}\psi )\gamma _{[a}\gamma _b\gamma _{c]} + 
\frac{mc}\hbar \right) \psi 
&=&0,  
\label{5} \\
R_{AB}-\frac {1}{2}G_{AB}R = 8l_{Pl}^2T_{AB}^D + 
8 l_{Pl}^4G_{AB}(\overline{\psi }
\gamma ^{[a}\gamma ^b\gamma ^{c]}\psi )
(\overline{\psi }\gamma _{[a}\gamma
_b\gamma _{c]}\psi ), &&  
\label{6}
\end{eqnarray}
here $S^{abc}$ is the antisymmetric torsion tensor, 
$l_{Pl}^2=\pi\hbar G/c^3$ ($l_{Pl}$ is Planck length), 
$R_{AB}$ is the 5D Ricci tensor. 
The stress-energy tensor $T_{AB}^D$ of the Dirac field is:  
\begin{equation} 
T_{AB}^D=-i\left[ \overline{\psi }
\gamma _A(\stackrel{\{\}}{\nabla }_B\psi
) + \overline{\psi} \gamma _B
(\stackrel{\{\}}{\nabla }_A\psi )\right]
+(Hermitian-conjugate),  
\label{7}
\end{equation} 
\setlength{\baselineskip}{3em}
here $\stackrel{\{\}}{\nabla }_A$ 
means the covariant derivative without
torsion. The 5D metric is defined in the ordinary way 
\setlength{\baselineskip}{2em} 
$G_{AB}=h_{\bullet B}^ah_{aB}$. 
We suppose that in this spacetime there are small regions 
where torsion plays the key role. 
In this case equation 
(\ref{5}) is the key equation for 
understanding what happens in such 
small regions. In this paper we want to demonstrate 
that the 5D Heisenberg equation, 
taking into account torsion, has solutions 
which are regular in the whole spacetime 
where the gravitational effect comes only from torsion. 
\par 
The ansatz for equation (\ref{5}) is taken as the 
standard spherically symmetric spinor: 
\begin{equation}
\psi (r,t)=e^{i\omega t}\left\{ 
\begin{array}{c}
f(r) \\ 
0 \\ 
ig(r)\cos \theta  \\ 
ig(r)\sin \theta e^{i\varphi }
\end{array}
\right\} ,  
\label{9}
\end{equation}
here $r,\theta ,\varphi $ are the spherical coordinates. 
It is interesting to note that in the 4D case the scalar 
$(\overline{\psi }\gamma 
^{[a}\gamma ^b\gamma ^{c]}\psi )
(\overline{\psi }\gamma _{[a}\gamma _b\gamma
_{c]}\psi )$ is not spherically symmetric. 
The nonlinear term 
$(\overline{\psi }\gamma^{[a}\gamma ^b\gamma ^{c]}\psi )
(\gamma _{[a}\gamma _b\gamma _{c]}\psi )$ 
in the Heisenberg equation is 
spherical symmetric only in 5D spacetime (see Appendix B). 
This is highly unusual and evidently points 
\textbf{\textit{to the close connection between 
torsion and multidimensional gravity.}} 
The substitution of Eq. (\ref{9}) into 
Heisenberg's Eq. (\ref{5}) gives us 
the following two equations (remember that we ignore 
the gravitational effects connected with the metric): 
\begin{eqnarray}
g^{\prime } + (-m + \omega )f + \frac{2g}r - 
12 l_{Pl}^2f\left( f^2-g^2\right)  & = & 0,
\label{10-1} \\
f^{\prime } - (m + \omega )g - 12 l_{Pl}^2g
\left( f^2-g^2\right) & = & 0  
\label{10-2}
\end{eqnarray}
here $\hbar ,c=1$. 
These equations coincide identically with equations 
for the 4D Heisenberg equation which have been 
investigated in Refs \cite{fin1}-\cite{fin2} with other
nonlinear terms 
($|\overline{\psi }\psi |^2$ and 
$|\overline{\psi }\gamma ^\mu\psi |^2$). 
Equations (\ref{10-1}) - (\ref{10-2}) have regular solutions 
in all space only for some discrete set of initial 
values $f(0)$ and $g(0)$. Near the origin the regular 
solution has the following behavior: 
\begin{eqnarray}
g(r) = g_1r + g_3\frac{r^3}{6} + \cdots ,
\label{1e}\\
f(r) = f_0 + f_2\frac{r^2}{2} + \cdots 
\label{1f}
\end{eqnarray}
The substitution (\ref{1e}) -(\ref{1f}) into 
Eqs (\ref{10-1}) - (\ref{10-2}) give 
us: 
\begin{eqnarray}
g_1 = \frac{f_0}{3}\left [12l^2_{Pl} f^2_0 + (m - \omega) \right ], 
\label{1g}\\
f_2 = g_1\left [12l^2_{Pl} f^2_0 + (m + \omega) \right ]. 
\label{1h}
\end{eqnarray}
This means that for fixed $m$ and $\omega$ a solution depends 
only on the initial value $f(0) = f_0$.  For arbitrary $f_0$ 
values the solution is singular at infinity ($r\to \infty$). 
However there are a discrete series of $(f_0)_n$ values for which 
the solutions become regular at infinity (n is the number of 
intersections that $f(r)$ ($g(r)$) make with the $r$-axe). 
Each of these solutions 
has finite mass and spin. The ground state is for $n=0$. More 
detailed discussion about properties of these solutions 
can be found in Ref. \cite{fin1}, \cite{fin2}. 
Thus, Eqs (\ref{10-1}) and (\ref{10-2}) have a discrete 
spectrum of regular solutions in all space, and 
they have finite energy. 
At infinity $(r\to\infty)$ these 
solutions have the following asymptotical behaviour: 
\begin{eqnarray} 
f = f_\infty + \frac{ae^{-\alpha r}}{r^2} + \cdots ,
\label{11-1}\\
g = \frac{be^{-\alpha r}}{r^2} + \cdots ,
\label{11-2}\\
\alpha ^2 = 4\omega (m + \omega),
\label{11-3}\\
\frac{b}{a} = -\sqrt{1 - \frac{m}{\omega}}, 
\label{11-4}\\ 
f_\infty = \pm \sqrt{\frac{-m + \omega}{12 l^2_{Pl}}}. 
\label{11-5} 
\end{eqnarray}
This guarantees the finiteness of mass, energy and so on.
If the energy of such solutions is 
$\omega \approx E_{Pl}= (\hbar c^5/G)^{1/2}$ 
and $m=0$ then these solutions are essentially nonzero 
only in the Planck region. 

\section{Discussion}

{\bf 1.} We see that the linear size 
of these solutions is on the order of the Planck length. 
The energy, spin and action 
are finite for these solutions and they are concentrated 
in the Planck region. This means that these 
solutions will give an essential 
contribution to the Feynman path integral in quantum 
gravity. All these 
results allow us to hypothesize that 
torsion can play a very important 
role in the formation of bubbles in the 
spacetime foam and in preventing 
cosmological singularities \cite{kop}, \cite{gas}. 
In conclusion I would like to underline that 
probably \textbf{\textit{torsion + multidimensional gravity}} 
is a more natural object in nature than either 
object separately. 
\par 
{\bf 2.} It is well known that boson (gauge) fields 
have a geometrical interpretation in multidimensional 
Kaluza-Klein theories as off-diagonal components 
of a multidimensional metric. Unfortunately the 
fermion fields do not have a similar geometrical 
interpretation. This is a big obstacle to 
Einstein's point of view that nature is 
\textbf{\textit{pure geometry}}. It is possible that 
Eq. (\ref{4}) can be considered as an 
algebraic relation for geometrization of the classical 
spinor field. Let us examine it not in terms of the 
right side (spinor field) as the source for the left 
side (torsion) but \textit{vice versa}: torsion $S^{abc}$ is 
the source 
of the classical spinor field $\psi$. In this 
case we can say that the classical spinor field 
in some sense is ``the square root'' of torsion 
just the Dirac equation is ``the square root'' 
of the Klein-Gordon equation. Such a point of view 
immediately leads to the conclusion that 
\textit{\textbf{\underline{nonpropagating torsion} 
in Einstein-Cartan gravity is the geometrical 
source for fermion fields.}} In 4D gravity Eq. (\ref{4}) 
can be written as 
$S^\mu \propto \overline{\psi }\gamma ^\mu 
\gamma ^5 \psi$ where in the left and right sides 
of the equation we have 4 number of independent components. 
In order to define the 4 spinor components we have 
4 quadratic algebraic equations. In this case 
the Heisenberg equation (\ref{5}) is the gravity equation! 
In the 5D case we can introduce 
the tensor $\Sigma^{ab}$ instead of $S^{abc}$: 
$\Sigma^{ab} = E^{abcde}S_{cde}$ (here $E^{abcde}$ 
is the 5D completely antisymmetrical Levi-Civita 
tensor). The antisymmetrical tensor $\Sigma^{ab}$ 
has the $\frac{(5\times 5)-5}{2}=10$ number of 
independent components. Hence each component of the spinor 
would be some nonlinear combination of 10 torsion 
components. But in fact the Heisenberg equation 
(\ref{5}) has only 4 components, hence for 
geometrization of the spinor field in 5D space we 
should have torsion with some algebraic 
restrictions on torsion.

\section{Acknowledgments}

This project has been funded by the National Research 
Council under the Collaboration in Basic Science 
and Engineering Program.
I would also like to 
thank D.Singleton for his invitation to research at VCU.

\appendix

\section{Riemann-Cartan geometry}

Here we give the necessary notions of Riemann-Cartan 
geometry as in Ref. \cite{hehl}. The affine connection of 
Riemann-Cartan spacetime is: 
\begin{equation}
\Gamma^A_{\bullet BC} =  \left \{ \,^B_{AC}\right \}
+ {S_{BC}}^A_\bullet - {S_C}^A_{\bullet B} + S^A_{\bullet AB},  
\label{12}
\end{equation}
here $\left \{ \,^B_{AC}\right \}$ are Christoffel symbols. 
Cartan's torsion tensor ${S_{BC}}^A_.$ is defined 
according to : 
\begin{equation}
{S_{BC}}^A_\bullet = S^A_{\bullet BC} = 
\frac{1}{2}\Gamma ^A_{[BC]} = 
\frac{1}{2} \left (\Gamma^A_{\bullet BC} - \Gamma^A_{\bullet CB} \right ). 
\label{13}
\end{equation}
The contorsion tensor is: 
\begin{equation}
{K_{BC}}^A_\bullet = {S_{BC}}^A_\bullet + {S_C}^A_{\bullet B} - 
S^A_{\bullet AB}
\label{14}
\end{equation}
In this case the affine connection is: 
\begin{equation}
\Gamma^A_{\bullet BC} = \left \{ \,^B_{AC}\right \} 
- {K_{BC}}^A_\bullet .
\label{15}
\end{equation}
The Riemann curvature tensor is defined in the usual 
way as: 
\begin{eqnarray} 
R^A_{\bullet BCD} = \partial_C\Gamma ^A_{\bullet BD} - 
\partial_D\Gamma ^A_{\bullet BC} + 
\Gamma ^A_{\bullet EC}\Gamma ^E_{\bullet BD} - 
\Gamma ^A_{\bullet ED}\Gamma ^E_{\bullet BC} = 
\nonumber \\
\stackrel{\{\}}{R^A}_{\bullet BCD} + \nabla _D K^A_{\bullet BC} - 
\nabla _C K^A_{\bullet BD} + K^A_{\bullet EC}K^E_{\bullet BD} - 
K^A_{\bullet ED}K^E_{\bullet BC}
\label{16}
\end{eqnarray}
A modified torsion tensor is: 
\begin{equation}
{T_{BC}}^A_\bullet = {S_{BC}}^A_\bullet + 
\delta ^A_B {S_{CD}}^D_\bullet - 
\delta ^A_C {S_{BD}}^D_\bullet .
\label{17}
\end{equation}
We can decompose the curvature scalar into Riemannian 
and contorsion pieces as follows: 
\begin{equation}
R = \stackrel{\{\}}{R} +  
2\stackrel{\{\}}{\nabla}_A \left ({K_B^\bullet }^{AB} \right )- 
{T_A^\bullet }^{BC}{K_{CB}}^A_\bullet 
\label{18}
\end{equation}
For antisymmetric torsion we can write: 
\begin{eqnarray}
S_{ABC} = T_{ABC}; \qquad K_{ABC} = - S_{ABC},
\label{19-1}\\
R = \stackrel{\{\}}{R} - S_{ABC} S^{ABC}.
\label{19-2}
\end{eqnarray}

\section{Cubic term in 4D and 5D Heisenberg equations}

The calculation of 
$\left (\overline{\psi }\gamma^{[\alpha}
\gamma ^\beta\gamma ^{\delta]}\psi \right )
\left (\gamma _{[\alpha}\gamma _\beta\gamma _{\delta]}\psi \right )$ 
in 4D spacetime ($\alpha ,\beta , \delta = 0,1,2,3$) give us the 
following result: 
\begin{eqnarray}
\left (\overline{\psi }\gamma^{[\alpha}\gamma 
^\beta\gamma ^{\delta]}\psi \right )
\left (\gamma _{[\alpha}\gamma _\beta
\gamma _{\delta]}\psi \right ) 
\propto 
e^{i\omega t}\left\{ 
\begin{array}{c}
6f(f^2 - 2g^2\sin ^2\theta + g^2) \\ 
12 \sin \theta \cos \theta fg^2 e^{i\varphi} \\ 
-6i\cos \theta g (f^2 + g^2)\\ 
6i\sin \theta g (f^2 - g^2) e^{i\varphi }
\end{array}
\right\} ,  
\label{b1}\\
\left (\overline{\psi }\gamma^{[\alpha}
\gamma ^\beta\gamma ^{\delta]}\psi \right )
\left (\overline{\psi }\gamma _{[\alpha}
\gamma _\beta\gamma _{\delta]}\psi \right ) 
\propto 
\left (f^4 - 4f^2g^2\sin ^2\theta + 2f^2g^2 
+ g^4\right ) ,
\label{b2}
\end{eqnarray}
and we see that term (\ref{b1}) can not be included in Heisenberg 
equation (\ref{5}) as it is inconsistent with spinor ansatz 
(\ref{9}). In 5D spacetime the analogous cubic term 
$\left (\overline{\psi }\gamma^{[a}
\gamma ^b\gamma ^{c]}\psi \right )
\left (\gamma _{[a}\gamma _b\gamma _{c]}\psi \right )$ 
has the following form: 
\begin{eqnarray}
\left (\overline{\psi }\gamma^{[a}
\gamma ^b\gamma ^{c]}\psi \right )
\left (\gamma _{[a}\gamma _b\gamma _{c]}\psi \right ) 
\propto 
12 e^{i\omega t}\left (f^2 - g^2 \right )
\left\{ 
\begin{array}{c}
f \\ 
0 \\ 
ig\cos \theta  \\ 
ig\sin \theta e^{i\varphi }
\end{array}
\right\} ,  
\label{b3}\\
\left (\overline{\psi }\gamma^{[a}\gamma ^b
\gamma ^{c]}\psi \right )
\left (\overline{\psi }\gamma _{[a}\gamma _b
\gamma _{c]}\psi \right ) 
\propto 
\left (f^2 - g^2\right )^2
\label{b4}
\end{eqnarray}
and the dependance on the polar angles $\theta$ and $\varphi$ 
in (\ref{b3}) is the same as in the ansatz (\ref{9}).

\end{document}